\documentclass[aps,pra,reprint,superscriptaddress]{revtex4-1}
\usepackage{amsmath}
\usepackage{graphicx}
\usepackage{natbib}
\usepackage{color}
\usepackage{bm}
\newcommand{\bee}{\begin{equation}}
\newcommand{\ene}{\end{equation}}

\newcommand{\h}[1]{\widehat{#1}}
\renewcommand{\phi}{\varphi}
\newcommand{\s}{\h{\sigma}}
\newcommand{\om}{\omega}

\newcommand{\Cr}{\h{a}^{+}}
\newcommand{\An}{\h{a}}

\begin{document}
\title{Decoupling light and matter: permanent dipole moment induced collapse of Rabi oscillations}

\author{Denis G. Baranov}
\email[]{denisb@chalmers.se}
\affiliation{Department of Physics, Chalmers University of Technology, 412 96 Gothenburg, Sweden}
\affiliation{Moscow Institute of Physics and Technology, 9 Institutskiy per., Dolgoprudny 141700, Russia}

\author{Mihail I. Petrov}
\affiliation{ITMO University, St. Petersburg 197101, Russia}

\author{Alexander E. Krasnok}
\affiliation{ITMO University, St. Petersburg 197101, Russia}
\affiliation{Department of Electrical and Computer Engineering, The University of Texas at Austin, Austin, Texas 78712, USA}

\date{\today}

\begin{abstract}
Rabi oscillations is a key phenomenon among the variety of quantum optical effects that manifests itself in the periodic oscillations of a two-level system between the ground and excited states when interacting with electromagnetic field. Commonly, the rate of these oscillations  scales proportionally with the magnitude of the electric field probed by the two-level system. Here, we investigate the interaction of light with a two-level quantum emitter possessing permanent dipole moments. The  semi-classical approach to this problem predicts  slowing down and even full suppression of Rabi oscillations due to asymmetry in diagonal components of the dipole moment operator of the two-level system. We consider behavior of the system in the fully quantized picture and  establish the analytical condition of Rabi oscillations collapse. These results for the first time emphasize the behavior of two-level systems with permanent dipole moments in the few photon regime, and suggest observation of novel quantum optical effects.
\end{abstract}

\maketitle

\section{Introduction}
Theory of a two-level system (TLS) interacting with electromagnetic field is of prime importance for a wide spectrum of applied problems, including laser science \cite{Sargent,Scully}, fluorescent spectroscopy~\cite{Lakowicz2008}, nano-imaging~\cite{SandoghdarN2000, Acuna2012,  Beams2013}, design of single photon sources~\cite{Arcari2014, Lodahl2015, Rao2007, Hoang2016} and efficient light emitting devices~\cite{Lodahl2015,LEDs}. It also plays the central role in the quantum information theory in the context of coherent qubits control~\cite{Chiorescu2003, Neeley2008, Wallraff2004, Lupascu2007}. Due to its very general formalism, it is equally applicable to the wide range of different electronic systems: electric and magnetic atomic and molecular transitions, quantum dots and quantum wells, superconducting Josephson junctions, defect centers in nanocrystals and others. 

In the theoretical description of light-matter interaction, it is often assumed that dynamics of the system is governed by the non-diagonal matrix element of the dipole moment operator. However, certain systems possess non-zero permanent dipole moments (PDM). An example of such system is a polar molecule, an atom polarized by static electric field~\cite{Potassium} or an asymmetric quantum dot~\cite{Savenko}. Magnetic dipole atomic transitions, e.g., in rare-earth ions of Eu$^{3+}$~\cite{Zia2012,Carminati2014,Novotny2015} may also have non-zero permanent magnetic dipoles, in contrast to the case of electric dipole transitions, where diagonal elements of the electric dipole moment operator are \emph{always zero} for atomic eigenstates \cite{Landau}, what follows from the parity of the wave function. 

Quantum systems with electric PDM have been widely investigated in the context of multi-photon processes~\cite{Meath84,Meath85,Band90,Meath92, AIP}. It was shown that presence of permanent dipole moment modifies multi-photon absorption rates. Emission spectrum features of quantum systems possessing permanent dipole moments were studied in Refs.~\cite{Hoffmann,Savenko}, where it was shown that such a system can radiate at Rabi frequency and serve as an emitter in the THz range. More recently, it was demonstrated that a two-level system with permanent dipoles can be inversely populated in the steady state if acted by two monochromatic fields~\cite{Macovei}. 

\begin{figure}[!b] 
\includegraphics[width=0.45\textwidth]{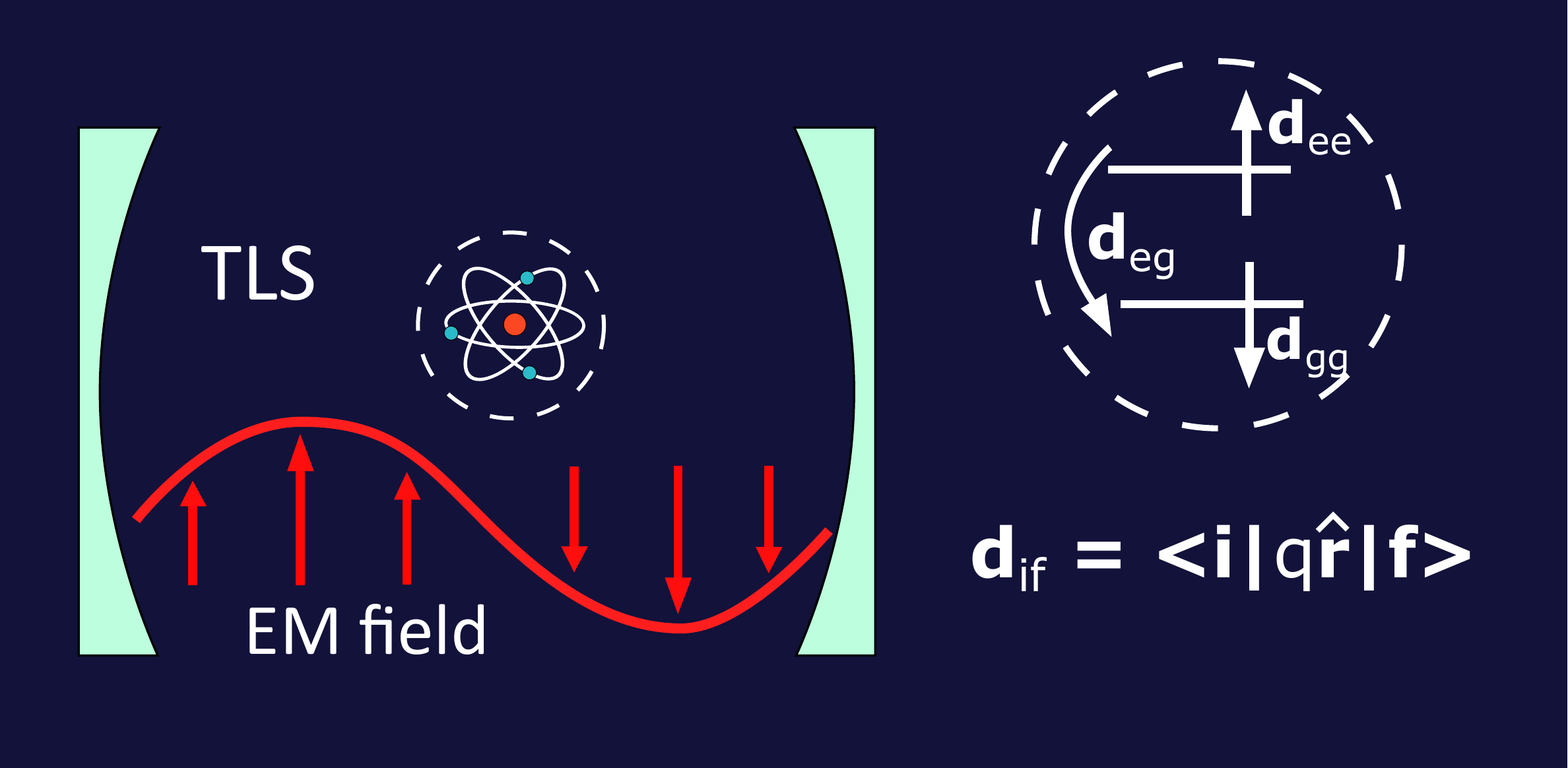}
\caption{\label{fig1}A schematic of the system under study. A two-level system modeling a quantum emitter interacts with electromagnetic field of a cavity resulting in Rabi oscillations of the two-level system population inversion. Inset on the right shows the internal structure of the two-level system with permanent dipole moments.}
\end{figure}

Here, we investigate in more details the Rabi oscillations of a TLS with PDM.
One of the main effects in such systems is that PDM enables multi-photon Rabi oscillations and strongly affects the Rabi frequency for single photon process.
Since the pioneering paper of Meth et. al.~\cite{Meath84}, it was shown that the Rabi oscillations in principle may be suppressed by tuning the magnitude of the emitter PDM. 
However, the treatment of this problem has been mostly limited to the semiclassical approach.
In this work, we address the fully quantum picture of interaction of light with a permanent dipole TLS.
By means of numerical simulations and theoretical analysis, we find new eigenstates of the PDM TLS--single mode cavity Hamiltonian and observe the collapse of Rabi oscillations in the few photon regime.
Our findings demonstrate previously unexplored regime of light-matter interaction and provide novel tools for coherent control of quantum emitters possessing PDM.

\section{Semiclassical description}

We start from a brief overview of the semiclassical description of the dynamics of a TLS with PDM in the presence of light field.  
The system under study is schematically depicted in Fig. 1. It consists of a generalized TLS interacting with electromagnetic field. The TLS can make transition between the ground $\left| g \right\rangle $ and excited $\left| e \right\rangle $ states, respectively, separated by the energy $\hbar \omega_0$. 

To begin with, we consider the simplest scenario of a TLS driven by a classical monochromatic electromagnetic wave ${\bf{E}}\left( t \right) = {{\bf{E}}_0}\cos \left( {{\omega _0}t} \right)$. In what follows we will use the notation ${\bf{E}}\left( t \right)$ for description of the field component interacting with the TLS, implying the electric field. However, this formalism is equally applicable to any other TLS interacting with either electric or magnetic oscillating field and therefore this does not limit the ubiquity of our results. Despite simplicity of this model, it captures important signatures in the dynamics of a TLS with PDM. The Hamiltonian of the system has the standard form $\hat H = {\hat H_{{\rm{TLS}}}} + {\hat H_{{\mathop{\rm int}} }}$, where the TLS part is $\hat H = \hbar {\omega _0}{{\hat \sigma }^\dag }\hat \sigma$ and the interaction part is ${\hat H_{{\mathop{\rm int}} }} =  - {\bf{E}}\left( t \right){\bf{\hat d}}$. The dipole moment operator ${\bf{\hat d}}$ has the form
\begin{equation}
{\bf{\hat d}} = {{\bf{d}}_{ge}}\hat \sigma  + {\bf{d}}_{eg}{\hat \sigma ^\dag } + {{\bf{d}}_{gg}}\hat \sigma {\hat \sigma ^\dag } + {{\bf{d}}_{ee}}{\hat \sigma ^\dag }\hat \sigma,
\label{eq1}
\end{equation}
Here $\hat \sigma = \left| g \right\rangle \left\langle e \right|$ is the lowering operator, ${{\bf{d}}_{eg}}={\bf{d}}_{ge}^{*}$ is the transition dipole moment, and ${{\bf{d}}_{ee}}$ and ${{\bf{d}}_{gg}}$ are the permanent dipole moments which are equal to zero in the common Jaynes-Cummings model \cite{Scully}. Without loss of generality we will assume the that ${\bf{d}}_{eg}$ along with ${{\bf{d}}_{ee}}$ and ${\bf d}_{gg}$ are all real-valued.

 With the use of expression~(\ref{eq1}) for the dipole moment operator we write the Hamiltonian in the symmetric form:
\begin{equation}
\begin{gathered}
\hat H =  - \frac{{\hbar {\omega _0}}}{2}\hat \sigma {{\hat \sigma }^\dag } + \frac{{\hbar {\omega _0}}}{2}{{\hat \sigma }^\dag }\hat \sigma  - \hbar {\Omega _R}\cos \left( {{\omega _0}t} \right){{\hat \sigma }^\dag }\hat \sigma \\
 - {{\bf{d}}_{gg}}{{\bf{E}}_0}\cos \left( {{\omega _0}t} \right)\hat \sigma {{\hat \sigma }^\dag } - {{\bf{d}}_{ee}}{{\bf{E}}_0}\cos \left( {{\omega _0}t} \right){{\hat \sigma }^\dag }\hat \sigma 
\end{gathered} 
\end{equation}
where $\hbar {\Omega _R} = {{\bf{E}}_0}{{\bf{d}}_{eg}}$ is the Rabi frequency. In the absence of permanent dipoles, the population inversion of the TLS oscillates in time with frequency $\Omega_R$.

\begin{figure}[!t] 
\includegraphics[width=1\columnwidth]{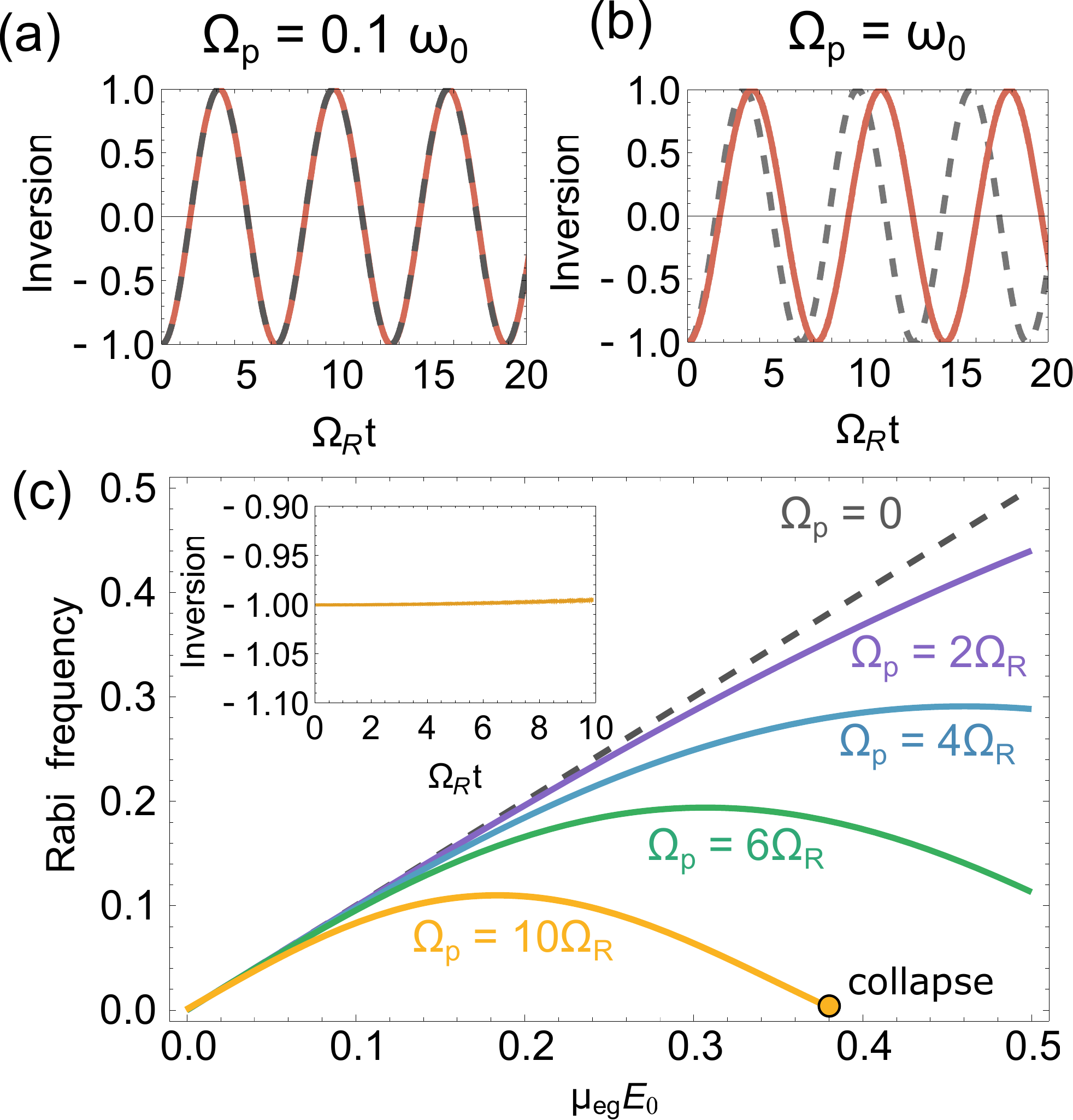}
\caption{(a, b) Population inversion of a TLS with non-zero permanent dipoles (solid) and an equivalent TLS without permanent dipoles (dashed) in a monochromatic electric field at different values of $\Omega_p$. Significant delay becomes visible in the permanent dipoles TLS dynamics. (c) Modified Rabi frequency of the TLS with permanent dipole moments for different $\Omega_p/\Omega_R$ as a function of applied electric field $E_0$. Dashed line shows unchanged Rabi frequency of a TLS without permanent dipoles. Inset: temporal dynamics of the population inversion at the collapse point corresponding to the first zero of the first order Bessel function $J_1$, $\Omega_{p}\approx 3.8\omega_0$.}
\label{fig2}
\end{figure}

Representing the TLS wave function in the general form $\left| {\Psi \left( t \right)} \right\rangle  = {c_e}\left( t \right)\left| e \right\rangle  + {c_g}\left( t \right)\left| g \right\rangle $ and substituting into the Schroedinger equation, we arrive at the following equations of motion describing temporal evolution of the TLS in the interaction representation:
\begin{equation} 
\begin{gathered}
 {{\dot c}_g}(t) = i \Omega_{R}\cos(\omega t)\exp\left(-i\left(\omega_0 t - \dfrac{\Omega_{p}}{\omega} \sin(\omega t)\right)\right)c_{e}(t), \\
 {{\dot c}_e}(t) = i \Omega_{R}\cos(\omega t)\exp\left(i\left(\omega_0 t - \dfrac{\Omega_{p}}{\omega} \sin(\omega t)\right)\right)c_{g}(t).
\end{gathered}
\label{eq3}
\end{equation}
Here we have introduced  the quantity $\hbar {\Omega _{{\text{p}}}} = {{\bf{E}}_0}\left( {{{\bf{d}}_{ee}} - {{\bf{d}}_{gg}}} \right)$, which will be referred to as the \emph{permanent} coupling constant. It shows the strength  of asymmetry between the dipole moment elements in ground and excited states.

Figures~\ref{fig2}(a),(b) show the TLS population inversion $W(t)={\left| {c_e\left( t \right)} \right|^2} - {\left| {c_g\left( t \right)} \right|^2}$ as a function of time obtained from the numerical solution of Eq.~\ref{eq3} for different ratios 
$\Omega_{p}/\omega_0$.
For comparison we also plot the population inversion of a pure non-diagonal system with ${{{\bf{d}}_{ee}} - {{\bf{d}}_{gg}}}=0$. 
The Rabi frequency is fixed at value $\Omega_R=0.01\omega_0$. The TLS is assumed to be initially in the ground state $\left| g \right\rangle $.
 It is seen that, in the regime of weak permanent coupling, ${\Omega _{p}} \ll {\omega_0}$, the TLS dynamics is unaffected by permanent dipoles [Fig. 2(a)]. 
However, in the strong permanent coupling regime, ${\Omega _{p}} \sim \omega_0$, the effect of permanent dipoles on the dynamics of the TLS becomes clearly visible, Fig. 2(b). Specifically, this effect consists in a larger period of Rabi oscillations cycle. 

The slowing of Rabi oscillations observed in the numerical simulations can be described by an elegant analytical expression derived in Refs.~\cite{Meath2000,Raikh,Hoffmann}. The resulting modified Rabi frequency is given by
\begin{equation} 
{\tilde \Omega _R} = 2{\omega _0}\frac{{{\Omega _R}}}{{{\Omega _{\text{p}}}}}{J_1}\left( {\frac{{{\Omega _{\text{p}}}}}{{{\omega _0}}}} \right),
\label{eq5}
\end{equation}
which yields $\Omega_R$ in the limit ${\Omega _{p}} \to 0$. The analytical estimation for the modified Rabi frequency $\tilde \Omega_R$ for emitters with different ratios of $\Omega_{\text{p}}/\Omega_R$ is presented in Fig. 2(c) as a function of applied oscillating electric field $E_0$. It clearly demonstrates the nonlinear dependence of the modified Rabi frequency shift with increasing $E_0$.

It immediately follows from Eq.~(\ref{eq5}) that the strength of the PDM effect is dictated by the relation between between the permanent coupling and the transition frequency.
It also shows that for observation of the predicted modification of inversion oscillations one does not need to achieve the ultrastrong coupling characterized by $\Omega_R \sim \omega_0$~\cite{Sirtori2010USC,Liberato2014}. Instead, it is sufficient to achieve only permanent ultrastrong coupling with $\Omega_{\text{p}} \sim \omega_0$ which is easier for quantum systems with large permanent dipoles $\left| {{{\bf{d}}_{ee}} - {{\bf{d}}_{gg}}} \right| \gg \left| {{{\bf{d}}_{eg}}} \right|$.

Equation~\ref{eq5} also suggests that the modified Rabi frequency may reach zero what would indicate \emph{collapse of Rabi oscillations}. The value of incident field $E_{\text{coll}}$ at which Rabi oscillations collapse occurs can be simply estimated by equating the argument of Bessel function to its first zero:
\begin{equation}
{E_{{\text{coll}}}} = {x_1}\frac{{\hbar {\omega _0}}}{{\left| {{{\mathbf{d}}_{ee}} - {{\mathbf{d}}_{gg}}} \right|}}
\label{eq6}
\end{equation}
with $x_1 \approx 3.8$ being the first zero of the $J_1$ Bessel function. The point of Rabi oscillations collapse can be found in Fig.~\ref{fig2}(c) for the TLS with $\Omega_{\text{p}}/\Omega_R=10$.
The corresponding temporal dynamics of the population inversion at the point of Rabi oscillations collapse $\hbar {\Omega _{{{p}}}} = {E_{{\rm{coll}}}}\left| {{{\bf{d}}_{ee}} - {{\bf{d}}_{gg}}} \right|$ is shown in the inset in Fig.~\ref{fig2}(c). Surprisingly, inversion of the TLS experiences only small deviations from its initial value $-1$ despite the considerable classical Rabi frequency $\Omega_R/\omega_0 \approx 0.38$. At this point we observe collapse of Rabi oscillations. With further increase of $\Omega_R$ and $\Omega_{p}$ Rabi oscillations recover, in consistence with expression~(\ref{eq5}).


The additional term in Hamiltonian (2) containing $\Omega_{p}$ describes fast modulation of the transition frequency $\omega_0$. From this point of view, the system can be treated as a parametric oscillator.
Given that the external electromagnetic field is tuned to the TLS transition frequency $\omega_0$, this modulation induced by the permanent dipoles introduces quickly variable detuning between the TLS and external field at certain time intervals.
A similar Rabi oscillations slowing has been reported recently for a spin-1/2 system in a slowly modulated longitudinal magnetic field~\cite{Raikh}.

The semiclassical approach presented above is commonly used for description of TLS dynamics with PDM as it draws a simple yet very illustrative picture. However, it does not allow one to understand the interaction of such TLS with single photons. To the best of our knowledge, only two works have treated the problem of PDM TLS interaction with light  within the fully quantum approach~\cite{Macovei, Savenko}. Nevertheless, the intriguing effect of Rabi oscillations collapse in the quantum regime has not been addressed.

\begin{figure*}[!t]
\includegraphics[width=.9\textwidth]{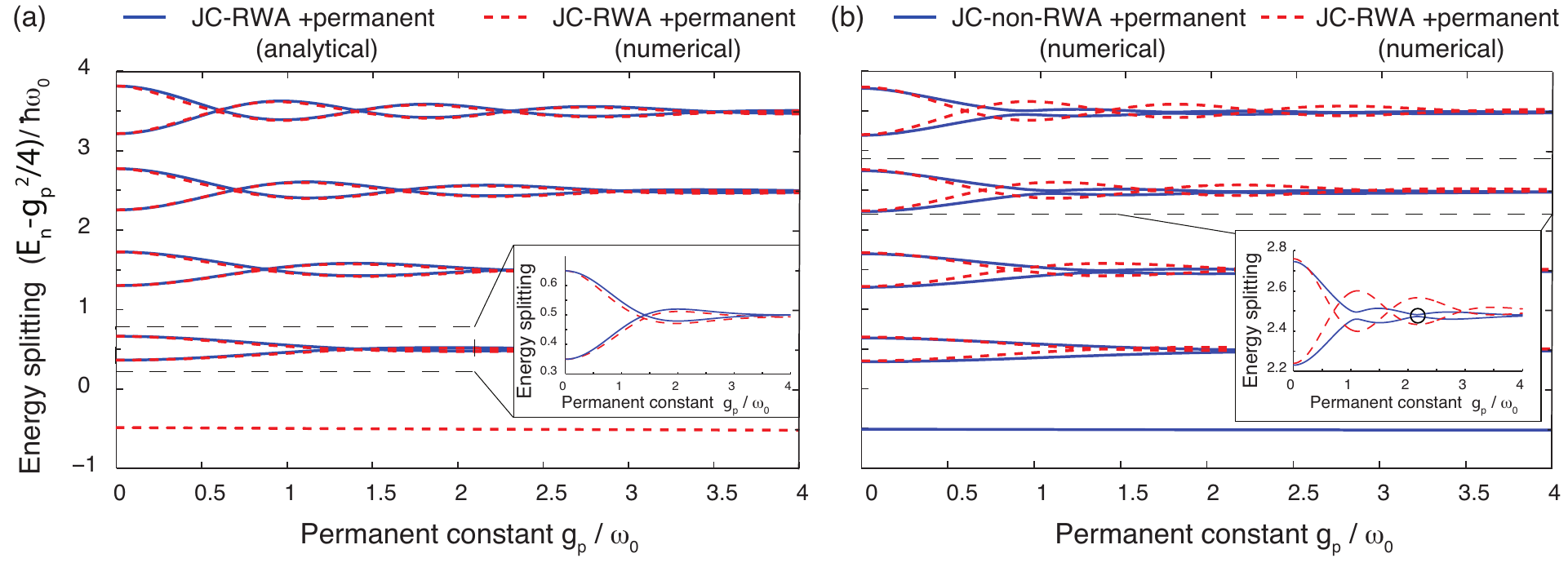}
\caption{The energy spectrum of the Hamiltonian~\ref{eq9} as a function of the permanent coupling constant $g_{p}$ with the JC part written in the RWA (a) and non-RWA (b) form. The solid line is the result of the exact numerical procedure, and the dashed line is obtained within  the perturbation theory. The JC coupling constant $g$ was taken be to $g/\omega_{0}=0.15$. The circle in the inset of panel (b) shows the point at which the Rabi oscillations are suppressed and dynamical equations are solved (see Fig.\ref{fig5}).}
\label{Permanent_JC}
\end{figure*} 


\section{Quantum description} 

We start our consideration with the standard fully quantized  Hamiltonian of a TLS interacting with a single electromagnetic mode:  
\begin{equation}
\hat H = \hbar {\omega _0}{\hat \sigma ^\dag }\hat \sigma  + \hbar \omega {\hat a^\dag }\hat a - {\mathbf{\hat d\hat E}}\left( {{{\mathbf{r}}_d}} \right)
\label{eq7}
\end{equation}
where $\omega$ is the resonator frequency, $\hat a$ is the annihilation operator of the resonator mode, ${\mathbf{\hat d}}$ is the dipole moment operator given by Eq. (1),  ${\bf{\hat E}}\left( {{{\bf{r}}_d}} \right) = {\cal E}_{0}\left( {{{\hat a}^\dag } + \hat a} \right)$  is the electric field operator at the TLS position $\mathbf{r}_{d}$, and ${\cal E }_{0}=\sqrt{2\pi\hbar \omega_0/V}$ is the vacuum field amplitude with $V$ being the mode volume.   

In order to draw an analytical picture of the interaction, we  find the energy spectrum of the full Hamiltonian~(\ref{eq7}). Following the semiclassical case, where the system \eqref{eq5} is written in the interaction picture, we rewrite the Hamiltonian using the unitary transformation:
\begin{equation}
\h{H}'=\h{D}(z)\h{H}\h{D}^{-1}(z), \qquad z=\dfrac{iz_0}{\om}e^{-\frac{i\om t}2}\sin(\om t),
\end{equation}
where $\hat{D}(z)=\exp(z \h a^{+} - z^{*}\h a )$ is the displacement operator \cite{Scully}, and $z_0=-\mathcal{E}_{0}({\bf d_{ee}} + {\bf d_{gg}})/\hbar$. The new Hamiltonian $\hat{H}'$ can be separated into the following three components: 
\begin{equation} 
\begin{gathered}
\h{H}'=\h H_{0}+ \h H_{JC}+\h H_{p},\\
\h H_{0}=\dfrac{\hbar\om_{0}}2\s_{z}+\hbar \om \h{b}^{+}\h{b},\\ 
\h H_{JC}=\hbar g(\h{b}^{+}+\h{b})(\s_{+}+\s_{-}),\\
\h H_{p}=\dfrac{\hbar g_{p}}{2}(\h{b}^{+}+\h{b})\s_{z},\\
\end{gathered}
\label{eq9}
\end{equation} 
where $\s_z=\s_+\s_--\s_-\s_+$, $g=-\mathcal{E}_{0}{\bf d_{eg}}/\hbar$ is the Jaynes-Cummings (JC) coupling constant, $g_{\rm p}=-\mathcal{E}_{0}({\bf d_{gg}} - {\bf d_{ee}})/\hbar$ is the coupling constant related to the PDM contribution, and $\h{b}=\An-z$ is the new cavity annihilation operator. In Eq.~(\ref{eq9}) the  component $\h H_{0}$ corresponds to the energy of non-interacting free atom and field, the second $\h H_{JC}$ corresponds to the JC part, and the last component $\h H_p$ is the energy operator related to the permanent dipole contribution. We will also use the notation $\h H_{JC}^{RWA}=\hbar g(\h{b}^{+}\s_{-}+\s_{+}\h{b})$  for the JC term written in the rotating wave approximation (RWA).

Now we have two terms in the full Hamiltonian $\h H'$ that modify the interaction-free atom-field system: the JC part $\h H_{JC}$ and the PDM part $\h H_{\rm p}$. 
This interaction leads to Rabi splitting, which manifests itself in Rabi oscillations of the TLS population inversion.
The frequency of these oscillations is directly related to the magnitude of this splitting. 
Thus, the collapse of the Rabi oscillations in the quantum picture is accompanied by zero Rabi splitting, i.e., degeneracy of the eigen states of the system.

A natural way to approach the eigen states problem seems to treat the PDM part of the Hamiltonian as a perturbation over the classical JC Hamiltonian $\h H_{0}+\h H_{JC}$, which solution can be obtained either in RWA approximation [35] or beyond that [36]. However, the effect of Rabi oscillations collapse is expected to appear in the regime of strong permanent coupling, thus, such perturbative approach would not be valid. We will start with considering the JC part as a perturbation over the permanently polarized atomic system described by the Hamiltonian $\h H_{0-p}=\h H_{0}+\h H_{p}$. The case of diagonal part as a perturbation will be also considered for illustrative purposes in the end of this section. In the following, to get a simple and clear picture of the effect of permanent contribution, we will focus on the resonant case, which implies $\omega=\omega_{0}$.


\subsection{Strong permanent coupling: the  collapse of Rabi splitting}
We start with finding the eigen states of  $\h H_{0-p}$ Hamiltonian, which is already diagonal in terms of the atomic states as contains the $\h\sigma_{z}$ atomic operator only. To diagonalize  the photonic part we introduce the generalized (or extended) \cite{DeOliveira1990,Philbin2014}  coherent states following the recent works on exact solvability of the Rabi model reported in Ref.~\cite{Braak, Chen2012}. Using simple Bogoluybov's transformation $\h A=\h b+g_{p}/2$ and $\h B=\h b-g_{p}/2$ one can get the diagonal form of the Hamiltonian
\begin{equation} 
\begin{gathered}
\h H_{0-p}^{'}=\dfrac{\hbar \om_{0}}2 +\hbar \om_{0}\left(\h A^{+}\h A -\dfrac{g^{2}}{4} \right)|e \rangle \langle e|\\ 
+\hbar \om_{0}\left(\h B^{+}\h B -\dfrac{g^{2}}{4} \right)|g \rangle \langle g|,
\end{gathered}
\label{H0p}
\end{equation} 
with the eigen functions
\begin{equation}
\label{eigH0p}
|\psi_{n,A}\rangle=|n\rangle_{A}|e\rangle \qquad \mbox{and }\quad |\psi_{n,B}\rangle=|n\rangle_{B}|g\rangle.
 \end{equation}
Here $|n\rangle_{A}$ and $|n\rangle_{B}$ are the eigen states (number of photon states) for $\h A^{+}\h A$  and $\h B^{+}\h B$ operators respectively, and  can  be considered also as generalized coherent states constructed from the displaced Fock's states:  
\begin{equation}
|n\rangle_{A}=\h D({g_{p}}/2)|n\rangle \qquad |n\rangle_{B}=\h D(-{g_{p}}/2)|n\rangle.
\end{equation}

One can see from the form of the Hamiltonian \eqref{H0p} that the energy levels corresponding to the $|\psi_{n-1,A}\rangle $ and $|\psi_{n,B}\rangle $ states are doubly degenerate and equal to $E_{n}=\hbar \om_{0}n+g_{p}^{2}/4$, and the energy is quadratic with respect to the permanent coupling constant.  Using the degenerate  perturbation method in the basis \eqref{eigH0p} we can find the energy splitting between the levels, which is  defined by the matrix elements  $\langle\psi_{n-1,A}|\h H_{JC}^{RWA}|\psi_{n,B}\rangle$. This Rabi splitting is related to the  displacement operator matrix elements in the Fock's basis: 
\begin{gather*}
\tilde{\Omega}_{R, n}=2\langle\psi_{n-1,A}|\h H_{JC}^{RWA}|\psi_{n,B}\rangle=\\ 
g\left(\sqrt{n+1}\langle n+1|\h D(-g_{p}/\omega_{0})|n+1\rangle-\right.\\ 
\left.\dfrac{g_{p}}{2\omega_{0}}\langle n+1|\h D(-g_{p}/\omega_{0})|n\rangle\right).
\end{gather*}

 The matrix elements of the displacement operator can be directly computed \cite{Tanas1992} that  gives the final expression containing Laguerre polynomial $L_{n}$:     
\begin{equation} 
\begin{gathered}
\label{QuantumSplit}
\tilde{\Omega}_{R, n}=
\dfrac{\Omega_{R, n}}{2}\exp\left(-\dfrac{g_{p}^{2}}{2\omega_{0}^{2}}\right)\left[L_{n+1}\left(\dfrac{g_{p}^{2}}{\omega_{0}^{2}}\right)\right. \\
\left.+L_{n}\left(\dfrac{g_{p}^{2}}{\omega_{0}^{2}}\right)\right], 
\end{gathered}
\end{equation}
where $\Omega_{R, n}=2g\sqrt{n+1}$ is Rabi splitting in the JC model.

The dependence of Rabi splitting on the permanent coupling constant $g_{p}$ is illustrated in Fig.~\ref{Permanent_JC}(a). The calculations were performed with the JC part written in the RWA. The nonzero value of the coupling constant $g$ gives  finite  splitting of the eigenstates at $g_{p}=0$. Increase of the permanent coupling  constant $g_p$ results in oscillatory behavior of the energy levels, clearly seen in Fig.~\ref{Permanent_JC}(a). At the intersection of two spectral branches the Rabi splitting vanishes, which is the trace  of the Rabi oscillations collapse.
This occurs for a discrete series of values of $g_{p}$, which are determined as the roots of the expression \eqref{QuantumSplit}. This expression is the analogue of the formula \eqref{eq5} for the fully quantized case. According to the properties of the Laguerre polynomials \cite{Arfken1972}, the number of Rabi collapse points is the same as the order of the energy band (see Fig.~\ref{Permanent_JC}).

The permanent dipole term $\h{H}_{p}$ is not energy conserving since it contains the terms $\An \s_{-}$ and $\Cr \s_{+}$, which rotate at high frequency. Thus, it is not fully consistent to consider the permanent dipole Hamiltonian without keeping the non-RWA terms in the JC term. This is the well-known limitation of the RWA model~\cite{Irish,Braak}.
In order to investigate the energy spectrum of the system beyond the RWA, we calculate the eigenvalues of the Hamiltonian~(\ref{eq9}) numerically, Fig.~\ref{Permanent_JC}(b). The non-RWA terms lead to the interaction of the neighboring energy terms and the  degeneracy in the points of Rabi collapse in the RWA model [see dashed line in Fig.~\ref{Permanent_JC} (b)] is lifted and non-zero splitting is observed. 

\begin{figure}[!t] 
\includegraphics[width=1\columnwidth]{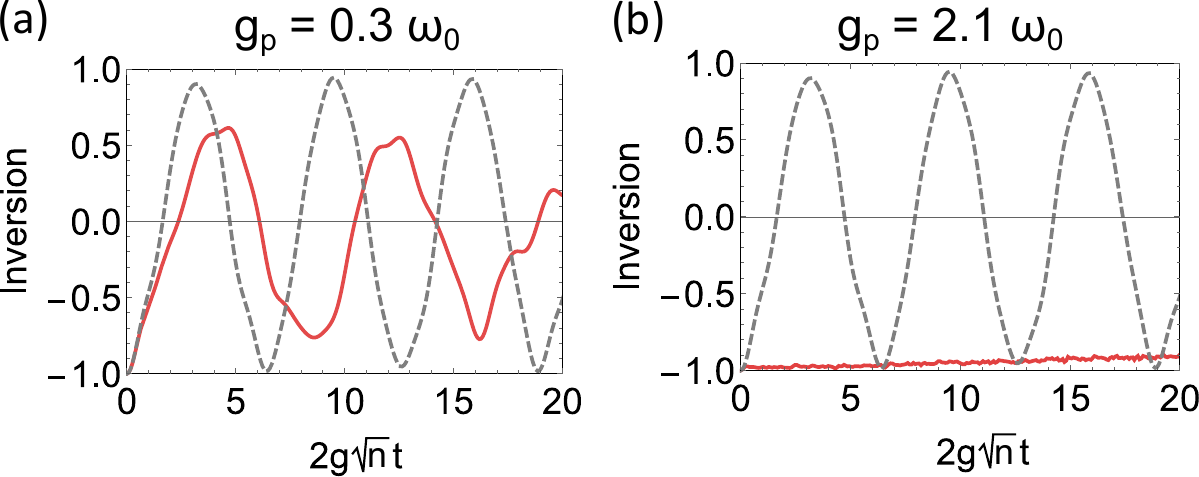}
\caption{Simulated population inversion dynamics of a TLS with non-zero permanent dipoles (solid) and an equivalent TLS without permanent dipoles (dashed) interacting with the three-photon Fock state of the cavity for different ratios $g_{p}/\omega_0$. Panel (b) corresponds to the intersection of the two spectral branches with vanishing Rabi splitting leading to collapse of the Rabi oscillations.}
\label{fig5}
\end{figure}

The predicted collapse of Rabi oscillations in the quantum regime is directly illustrated via numerical modeling of temporal dynamics of the system described by Hamiltonian~\eqref{eq9}. Dynamical behavior of the system initially placed in the Fock state $\left| {{\Psi _0}} \right\rangle  = \left| {g,n} \right\rangle$ is presented in Fig.~\ref{fig5}. The Schroedinger equations of motion in the non-RWA regime are solved for different ratios $g_p/\omega_0$ and the fixed cavity photons number $n=3$. The JC coupling constant is fixed at the value $g=0.15 \omega_0$ being consistent with Fig.~\ref{Permanent_JC}(b).
Similarly to the results of the semiclassical approach, frequency of the Rabi oscillations decreases with increasing $g_p$, Fig.~\ref{fig5}(a). However, in contrast to the semiclassical picture, not only the oscillations period is affected, but the amplitude of Rabi oscillations also changes [see Fig.~\ref{fig2}(b) for comparison]. Finally, at the intersection point of the two JC branches, shown in the inset of Fig.~\ref{Permanent_JC} (b) with circle, we observe collapse of Rabi oscillations, Fig.~\ref{fig5}(b).

\begin{figure*}[!t]
\includegraphics[width=.9\textwidth]{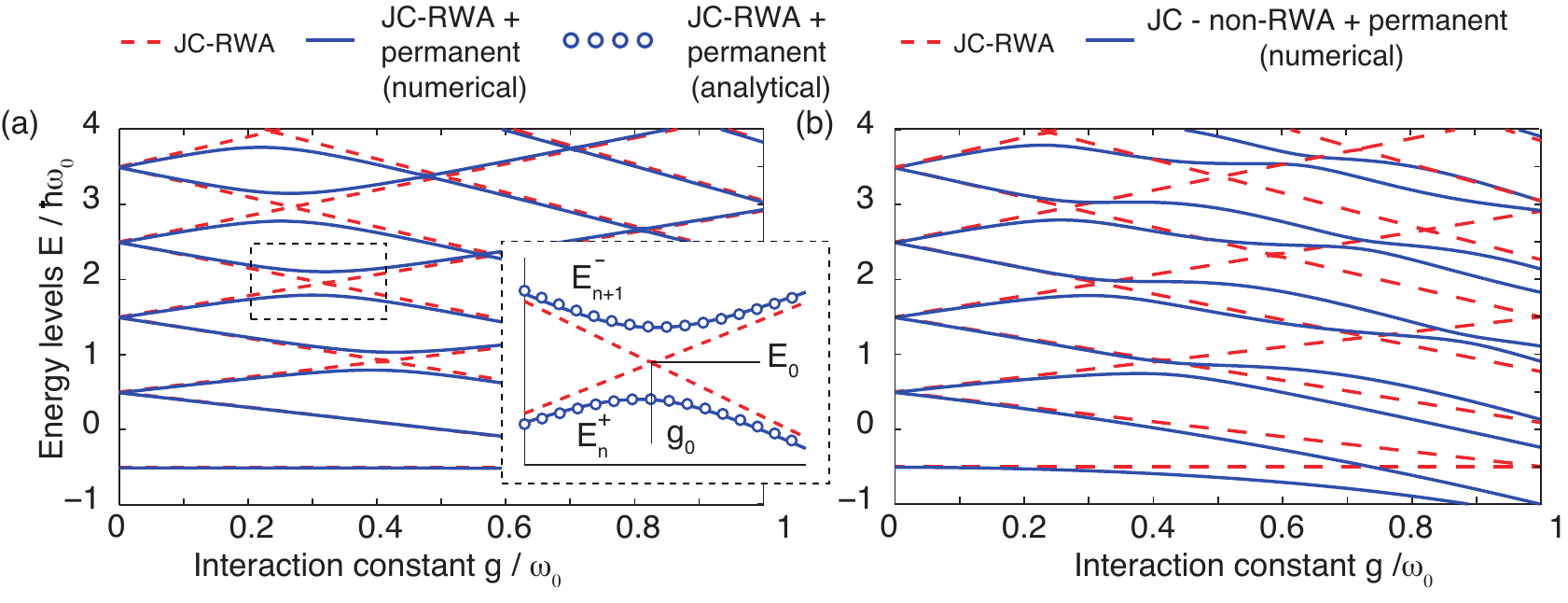}
\caption{(a) The energy levels of eigen states of the considered Hamiltonian as function of interaction constant  without permanent dipole ($g_{p}=0$, red dashed line) and with it ($g_{p}/\omega_0=0.1$, solid blue line). Inset:  perturbation theory results are shown with circles. (b) The energy of eigen states in non-RWA limit as function of interaction constant $g$ at fixed value of $g_p/\omega_0=0.1$. The red dashed line are the same as in (a)}
\label{JC_Permanent}
\end{figure*} 

\subsection{Strong non-diagonal coupling.}

In order to accomplish the picture of the eigen states of the system under study, we also consider  the case of strong non-diagonal coupling in more details, assuming that the diagonal part can be treated as a perturbation.  
Similar to the previous section, we  begin with  the JC Hamiltonian in the simplified RWA form, where the terms $\An \s_{-}$ and $\Cr \s_{+}$.
 The Eigenstates of the non-perturbed Hamiltonian $H_{JC}^{RWA}$ are the well-known dressed light-matter states \cite{Scully}, which in the case of resonant excitation $\om=\om_{0}$  have simple form~\cite{Meystre}:
\begin{equation}
\begin{gathered}
  \left| {\psi _n^ + } \right\rangle  = \frac{1}
{{\sqrt 2 }}\left| {g,n+1} \right\rangle  + \frac{1}
{{\sqrt 2 }}\left| {e,n } \right\rangle , \hfill \\
  \left| {\psi _{n }^ - } \right\rangle  = \frac{1}
{{\sqrt 2 }}\left| {e,n } \right\rangle  - \frac{1}
{{\sqrt 2 }}\left| {g,n+1} \right\rangle , \hfill \\ 
\end{gathered} 
\end{equation}
with the corresponding energy spectrum of the excited states given by
\begin{equation}
\begin{gathered}
  E_n^ \pm  = \left(n+\dfrac12\right)\hbar {\omega _0} \pm g\sqrt {n+1} ,n \ge 0 \hfill \\
\end{gathered} 
\end{equation}
The ground state has energy $ -{\hbar {\omega _0}}/2 $. The energy spectrum of the non-perturbed JC part is shown in Fig.~\ref{JC_Permanent}(a) and (b) with dashed red lines. Each line corresponds to a state with a fixed number of photons  $n$. Without interaction, $g=0$, the states with the same number of quanta are degenerate, but the energy splitting between them $E_{n}^{+}-E_{n}^{-}$ increases linearly with increase of $g$ (red dashed line). Thus, in the JC-RWA model only neighboring states   $\left| {\psi _n^ + } \right\rangle $ and $\left| {\psi _n^ - } \right\rangle $ interact. However, after introducing the permanent component $\h H_{p}$, which contains terms $\h b^{+}$ and $\h b^{-}$ changing the number of photons by 1, the neighboring polaritonic states $\left| {\psi _{n-1}^ + } \right\rangle $ and $\left| {\psi _{n}^ - } \right\rangle $ start to interact resulting in an anticrossing behavior (blue solid lines in Fig.~\ref{JC_Permanent}(a)).
The magnitude of the energy gap equals to approximately $(\sqrt{n+1}+\sqrt{n+2})g_{p}/2$, which can be found from the perturbation theory by  taking the $\left| {\psi _{n-1}^ + } \right\rangle $ and $\left| {\psi _{n}^ - } \right\rangle $ as non-perturbated solutions.

\begin{figure}[!t]
\includegraphics[width=1\columnwidth]{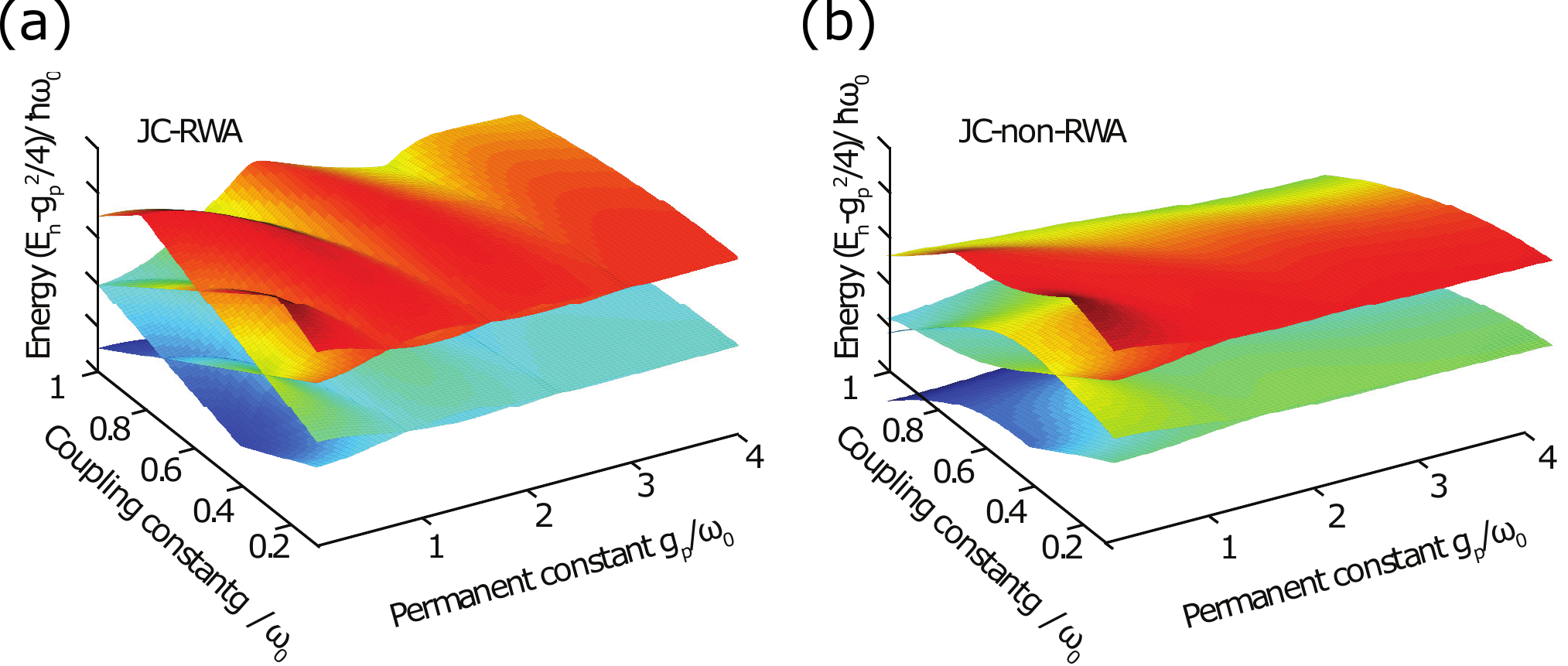}
\caption{ The two-dimensional plots of the eigen energies as the function of Permanent coupling constant  $g_{p}$ and JC coupling constant $g$ with JC part written in the RWA (a) and non-RWA (b) from. The four polaritonic branches with $n=2$ and $n=3$ are shown. }
\label{Permanent_JC_2D}
\end{figure}

Although the RWA allows us obtaining simple analytical results in the vicinity of energy terms crossings, the permanent Hamiltonian component contains non-RWA terms. In order to stay at the same level of generality we have  numerically computed the energy of the eigen states considering JC counterpart in non-RWA approximation, and the results are  shown in the Fig.~\ref{JC_Permanent} (b). Similar to the RWA case the gap opening for neighboring states still exists, but due to the higher order terms $\h b^{+} \h \sigma_{+}$ and $\h b \h \sigma_{-}$ in JC Hamiltonian,  the states differing in number of quanta $n$ by two also start to  interact, which was reported previously in Ref.~\cite{Irish}.

Finally, it is illustrative to summarize the analyzed cases in a 2D plot showing the eigen energy levels of the full Hamiltonian \eqref{eq9}  as a function of two coupling constants $g$ and $g_{p}$. This plot is show in Fig.~\ref{Permanent_JC_2D} for two cases of Jaynes-Cummings part written in RWA and non-RWA form. We have discussed in more details the cases of weak permanent and weak non-diagonal coupling, which correspond to cross section of the Fig.~\ref{Permanent_JC_2D} along $x$ and $y$ axis. One can also see the ripple-like behavior of the 2D surface in RWA approximation, which indicates that the collapse of Rabi splitting occurs not only for weak coupling regimes. However, this structure of the eigen states is smeared out in the non-RWA regime, shown in Fig.~\ref{Permanent_JC_2D} (b).        

\section{Outlook and Conslusion}
So far we have treated interaction of a TLS with a closed lossless cavity. At the same time, the Rabi oscillations slowing demonstrated here may have important effect on spontaneous decay of a TLS with permanent dipole. Indeed, in the weak coupling regime the spontaneous decay rate of a quantum emitter is given by $\Gamma  = 4\Omega _R^2/{\gamma _a}$ where $\gamma_a$ is the cavity mode decay rate~\cite{Scully}. In the particular case of a TLS with permanent dipoles Rabi frequency gets smaller with increasing $\Omega_p$. Thus we may expect that decay rate $\Gamma$ will also decrease for a TLS with permanent dipoles in comparison to a pure non-diagonal quantum emitter.

Another consequence of the predicted modification of Rabi oscillations may be found in the resonance florescence spectra of quantum emitters with permanent dipoles. In the simplest case, it is represented by the Mollow triplet~\cite{Scully} which consists of the central transition line and the two sidebands with frequencies $\omega_0 \pm \Omega_R$. Therefore, modification of $\Omega_R$ will shift the sidebands of the Mollow triplet.

To conclude, we have investigated the energy spectrum and temporal dynamics of a two-level quantum emitter with permanent dipole moments interacting with electromagnetic field. The presence of a permanent dipole moment has a crucial role on the Rabi oscillations of the quantum emitter. When the two-level system is driven by a classical field, Rabi oscillations frequency gradually decreases until oscillations completely collapse. We established the condition of Rabi oscillations collapse in semiclassical and quantum regime and discussed the implications of this phenomenon on other quantum optical effects.

\begin{acknowledgments}
Authors acknowledge fruitful discussion with Prof. O. V. Kibis. 
D.G.B. acknowledges support from RFBR project No 16-32-00444. M.I.P acknowledges support from RFBR project No 16-32-60167.
\end{acknowledgments}

\bibliography{tls}

\begin{thebibliography}{40}%
\makeatletter
\providecommand \@ifxundefined [1]{%
 \@ifx{#1\undefined}
}%
\providecommand \@ifnum [1]{%
 \ifnum #1\expandafter \@firstoftwo
 \else \expandafter \@secondoftwo
 \fi
}%
\providecommand \@ifx [1]{%
 \ifx #1\expandafter \@firstoftwo
 \else \expandafter \@secondoftwo
 \fi
}%
\providecommand \natexlab [1]{#1}%
\providecommand \enquote  [1]{``#1''}%
\providecommand \bibnamefont  [1]{#1}%
\providecommand \bibfnamefont [1]{#1}%
\providecommand \citenamefont [1]{#1}%
\providecommand \href@noop [0]{\@secondoftwo}%
\providecommand \href [0]{\begingroup \@sanitize@url \@href}%
\providecommand \@href[1]{\@@startlink{#1}\@@href}%
\providecommand \@@href[1]{\endgroup#1\@@endlink}%
\providecommand \@sanitize@url [0]{\catcode `\\12\catcode `\$12\catcode
  `\&12\catcode `\#12\catcode `\^12\catcode `\_12\catcode `\%12\relax}%
\providecommand \@@startlink[1]{}%
\providecommand \@@endlink[0]{}%
\providecommand \url  [0]{\begingroup\@sanitize@url \@url }%
\providecommand \@url [1]{\endgroup\@href {#1}{\urlprefix }}%
\providecommand \urlprefix  [0]{URL }%
\providecommand \Eprint [0]{\href }%
\providecommand \doibase [0]{http://dx.doi.org/}%
\providecommand \selectlanguage [0]{\@gobble}%
\providecommand \bibinfo  [0]{\@secondoftwo}%
\providecommand \bibfield  [0]{\@secondoftwo}%
\providecommand \translation [1]{[#1]}%
\providecommand \BibitemOpen [0]{}%
\providecommand \bibitemStop [0]{}%
\providecommand \bibitemNoStop [0]{.\EOS\space}%
\providecommand \EOS [0]{\spacefactor3000\relax}%
\providecommand \BibitemShut  [1]{\csname bibitem#1\endcsname}%
\let\auto@bib@innerbib\@empty
\bibitem [{\citenamefont {Sargent}\ \emph {et~al.}(1974)\citenamefont
  {Sargent}, \citenamefont {Scully},\ and\ \citenamefont {Lamb}}]{Sargent}%
  \BibitemOpen
  \bibfield  {author} {\bibinfo {author} {\bibfnamefont {M.}~\bibnamefont
  {Sargent}}, \bibinfo {author} {\bibfnamefont {M.~O.}\ \bibnamefont {Scully}},
  \ and\ \bibinfo {author} {\bibfnamefont {W.~E.}\ \bibnamefont {Lamb}},\
  }\href@noop {} {\emph {\bibinfo {title} {{Laser Physics}}}}\ (\bibinfo
  {publisher} {Addison-Wesley Pub. Co.},\ \bibinfo {year} {1974})\BibitemShut
  {NoStop}%
\bibitem [{\citenamefont {Scully}\ and\ \citenamefont
  {Zubairy}(1997)}]{Scully}%
  \BibitemOpen
  \bibfield  {author} {\bibinfo {author} {\bibfnamefont {M.~O.}\ \bibnamefont
  {Scully}}\ and\ \bibinfo {author} {\bibfnamefont {M.~S.}\ \bibnamefont
  {Zubairy}},\ }\href
  {http://www.amazon.com/Quantum-Optics-Marlan-O-Scully/dp/0521435951/ref=sr\_1\_6?s=books\&ie=UTF8\&qid=1398199694\&sr=1-6\&keywords=m+scully
  http://books.google.com/books?hl=en\&lr=\&id=20ISsQCKKmQC\&oi=fnd\&pg=PR19\&dq=Quantum+Optics\&ots=yRXTMSJCtq\&sig=pUkhft86lWP9mRRX9nGiA8HMKeI}
  {\emph {\bibinfo {title} {{Quantum optics}}}}\ (\bibinfo  {publisher}
  {Cambridge University Press},\ \bibinfo {address} {Cambridge},\ \bibinfo
  {year} {1997})\ p.\ \bibinfo {pages} {656}\BibitemShut {NoStop}%
\bibitem [{\citenamefont {Lakowicz}\ \emph {et~al.}(2008)\citenamefont
  {Lakowicz}, \citenamefont {Ray}, \citenamefont {Chowdhury}, \citenamefont
  {Szmacinski}, \citenamefont {Fu}, \citenamefont {Zhang},\ and\ \citenamefont
  {Nowaczyk}}]{Lakowicz2008}%
  \BibitemOpen
  \bibfield  {author} {\bibinfo {author} {\bibfnamefont {J.~R.}\ \bibnamefont
  {Lakowicz}}, \bibinfo {author} {\bibfnamefont {K.}~\bibnamefont {Ray}},
  \bibinfo {author} {\bibfnamefont {M.}~\bibnamefont {Chowdhury}}, \bibinfo
  {author} {\bibfnamefont {H.}~\bibnamefont {Szmacinski}}, \bibinfo {author}
  {\bibfnamefont {Y.}~\bibnamefont {Fu}}, \bibinfo {author} {\bibfnamefont
  {J.}~\bibnamefont {Zhang}}, \ and\ \bibinfo {author} {\bibfnamefont
  {K.}~\bibnamefont {Nowaczyk}},\ }\href {\doibase 10.1039/b802918k} {\bibfield
   {journal} {\bibinfo  {journal} {The Analyst}\ }\textbf {\bibinfo {volume}
  {133}},\ \bibinfo {pages} {1308} (\bibinfo {year} {2008})}\BibitemShut
  {NoStop}%
\bibitem [{\citenamefont {Mihaelis}\ \emph {et~al.}(2000)\citenamefont
  {Mihaelis}, \citenamefont {Hettich}, \citenamefont {Mlynek},\ and\
  \citenamefont {Sandoghdar}}]{SandoghdarN2000}%
  \BibitemOpen
  \bibfield  {author} {\bibinfo {author} {\bibfnamefont {J.}~\bibnamefont
  {Mihaelis}}, \bibinfo {author} {\bibfnamefont {C.}~\bibnamefont {Hettich}},
  \bibinfo {author} {\bibfnamefont {J.}~\bibnamefont {Mlynek}}, \ and\ \bibinfo
  {author} {\bibfnamefont {V.}~\bibnamefont {Sandoghdar}},\ }\href@noop {}
  {\bibfield  {journal} {\bibinfo  {journal} {Nature}\ }\textbf {\bibinfo
  {volume} {405}},\ \bibinfo {pages} {325} (\bibinfo {year}
  {2000})}\BibitemShut {NoStop}%
\bibitem [{\citenamefont {Acuna}\ \emph {et~al.}(2012)\citenamefont {Acuna},
  \citenamefont {M{\"{o}}ller}, \citenamefont {Holzmeister}, \citenamefont
  {Beater}, \citenamefont {Lalkens},\ and\ \citenamefont
  {Tinnefeld}}]{Acuna2012}%
  \BibitemOpen
  \bibfield  {author} {\bibinfo {author} {\bibfnamefont {G.~P.}\ \bibnamefont
  {Acuna}}, \bibinfo {author} {\bibfnamefont {F.~M.}\ \bibnamefont
  {M{\"{o}}ller}}, \bibinfo {author} {\bibfnamefont {P.}~\bibnamefont
  {Holzmeister}}, \bibinfo {author} {\bibfnamefont {S.}~\bibnamefont {Beater}},
  \bibinfo {author} {\bibfnamefont {B.}~\bibnamefont {Lalkens}}, \ and\
  \bibinfo {author} {\bibfnamefont {P.}~\bibnamefont {Tinnefeld}},\ }\href
  {\doibase 10.1126/science.1228638} {\bibfield  {journal} {\bibinfo  {journal}
  {Science}\ }\textbf {\bibinfo {volume} {338}},\ \bibinfo {pages} {506}
  (\bibinfo {year} {2012})}\BibitemShut {NoStop}%
\bibitem [{\citenamefont {Beams}\ \emph {et~al.}(2013)\citenamefont {Beams},
  \citenamefont {Smith}, \citenamefont {Johnson}, \citenamefont {Oh},
  \citenamefont {Novotny},\ and\ \citenamefont {Vamivakas}}]{Beams2013}%
  \BibitemOpen
  \bibfield  {author} {\bibinfo {author} {\bibfnamefont {R.}~\bibnamefont
  {Beams}}, \bibinfo {author} {\bibfnamefont {D.}~\bibnamefont {Smith}},
  \bibinfo {author} {\bibfnamefont {T.~W.}\ \bibnamefont {Johnson}}, \bibinfo
  {author} {\bibfnamefont {S.-H.}\ \bibnamefont {Oh}}, \bibinfo {author}
  {\bibfnamefont {L.}~\bibnamefont {Novotny}}, \ and\ \bibinfo {author}
  {\bibfnamefont {A.~N.}\ \bibnamefont {Vamivakas}},\ }\href {\doibase
  10.1021/nl401791v} {\bibfield  {journal} {\bibinfo  {journal} {Nano Lett.}\
  }\textbf {\bibinfo {volume} {13}},\ \bibinfo {pages} {3807} (\bibinfo {year}
  {2013})}\BibitemShut {NoStop}%
\bibitem [{\citenamefont {Arcari}\ \emph {et~al.}(2014)\citenamefont {Arcari},
  \citenamefont {S\"{o}llner}, \citenamefont {Javadi}, \citenamefont {Hansen},
  \citenamefont {Mahmoodian}, \citenamefont {Liu}, \citenamefont {Thyrrestrup},
  \citenamefont {Lee}, \citenamefont {Song}, \citenamefont {Stobbe},\ and\
  \citenamefont {Lodahl}}]{Arcari2014}%
  \BibitemOpen
  \bibfield  {author} {\bibinfo {author} {\bibfnamefont {M.}~\bibnamefont
  {Arcari}}, \bibinfo {author} {\bibfnamefont {I.}~\bibnamefont {S\"{o}llner}},
  \bibinfo {author} {\bibfnamefont {A.}~\bibnamefont {Javadi}}, \bibinfo
  {author} {\bibfnamefont {S.~L.}\ \bibnamefont {Hansen}}, \bibinfo {author}
  {\bibfnamefont {S.}~\bibnamefont {Mahmoodian}}, \bibinfo {author}
  {\bibfnamefont {J.}~\bibnamefont {Liu}}, \bibinfo {author} {\bibfnamefont
  {H.}~\bibnamefont {Thyrrestrup}}, \bibinfo {author} {\bibfnamefont {E.~H.}\
  \bibnamefont {Lee}}, \bibinfo {author} {\bibfnamefont {J.~D.}\ \bibnamefont
  {Song}}, \bibinfo {author} {\bibfnamefont {S.~R.}\ \bibnamefont {Stobbe}}, \
  and\ \bibinfo {author} {\bibfnamefont {P.}~\bibnamefont {Lodahl}},\
  }\href@noop {} {\bibfield  {journal} {\bibinfo  {journal} {Phys. Rev. Lett.}\
  }\textbf {\bibinfo {volume} {113}},\ \bibinfo {pages} {093603} (\bibinfo
  {year} {2014})}\BibitemShut {NoStop}%
\bibitem [{\citenamefont {Lodahl}\ \emph {et~al.}(2015)\citenamefont {Lodahl},
  \citenamefont {Mahmoodian},\ and\ \citenamefont {Stobbe}}]{Lodahl2015}%
  \BibitemOpen
  \bibfield  {author} {\bibinfo {author} {\bibfnamefont {P.}~\bibnamefont
  {Lodahl}}, \bibinfo {author} {\bibfnamefont {S.}~\bibnamefont {Mahmoodian}},
  \ and\ \bibinfo {author} {\bibfnamefont {S.}~\bibnamefont {Stobbe}},\
  }\href@noop {} {\bibfield  {journal} {\bibinfo  {journal} {Rev. Mod. Phys.}\
  }\textbf {\bibinfo {volume} {87}},\ \bibinfo {pages} {347} (\bibinfo {year}
  {2015})}\BibitemShut {NoStop}%
\bibitem [{\citenamefont {Rao}\ and\ \citenamefont {Hughes}(2007)}]{Rao2007}%
  \BibitemOpen
  \bibfield  {author} {\bibinfo {author} {\bibfnamefont {V.}~\bibnamefont
  {Rao}}\ and\ \bibinfo {author} {\bibfnamefont {S.}~\bibnamefont {Hughes}},\
  }\href@noop {} {\bibfield  {journal} {\bibinfo  {journal} {Phys. Rev. Lett.}\
  }\textbf {\bibinfo {volume} {99}},\ \bibinfo {pages} {193901} (\bibinfo
  {year} {2007})}\BibitemShut {NoStop}%
\bibitem [{\citenamefont {Hoang}\ \emph {et~al.}(2016)\citenamefont {Hoang},
  \citenamefont {Akselrod},\ and\ \citenamefont {Mikkelsen}}]{Hoang2016}%
  \BibitemOpen
  \bibfield  {author} {\bibinfo {author} {\bibfnamefont {T.~B.}\ \bibnamefont
  {Hoang}}, \bibinfo {author} {\bibfnamefont {G.~M.}\ \bibnamefont {Akselrod}},
  \ and\ \bibinfo {author} {\bibfnamefont {M.~H.}\ \bibnamefont {Mikkelsen}},\
  }\href {\doibase 10.1021/acs.nanolett.5b03724} {\bibfield  {journal}
  {\bibinfo  {journal} {Nano Lett.}\ }\textbf {\bibinfo {volume} {16}},\
  \bibinfo {pages} {270} (\bibinfo {year} {2016})}\BibitemShut {NoStop}%
\bibitem [{\citenamefont {Okamoto}\ \emph {et~al.}(2004)\citenamefont
  {Okamoto}, \citenamefont {Niki}, \citenamefont {Shvartser}, \citenamefont
  {Narukawa}, \citenamefont {Mukai},\ and\ \citenamefont {Scherer}}]{LEDs}%
  \BibitemOpen
  \bibfield  {author} {\bibinfo {author} {\bibfnamefont {K.}~\bibnamefont
  {Okamoto}}, \bibinfo {author} {\bibfnamefont {I.}~\bibnamefont {Niki}},
  \bibinfo {author} {\bibfnamefont {A.}~\bibnamefont {Shvartser}}, \bibinfo
  {author} {\bibfnamefont {Y.}~\bibnamefont {Narukawa}}, \bibinfo {author}
  {\bibfnamefont {T.}~\bibnamefont {Mukai}}, \ and\ \bibinfo {author}
  {\bibfnamefont {A.}~\bibnamefont {Scherer}},\ }\href@noop {} {\bibfield
  {journal} {\bibinfo  {journal} {Nat. Mater.}\ }\textbf {\bibinfo {volume}
  {3}},\ \bibinfo {pages} {601} (\bibinfo {year} {2004})}\BibitemShut {NoStop}%
\bibitem [{\citenamefont {Chiorescu}\ \emph {et~al.}(2003)\citenamefont
  {Chiorescu}, \citenamefont {Nakamura}, \citenamefont {Harmans},\ and\
  \citenamefont {Mooij}}]{Chiorescu2003}%
  \BibitemOpen
  \bibfield  {author} {\bibinfo {author} {\bibfnamefont {I.}~\bibnamefont
  {Chiorescu}}, \bibinfo {author} {\bibfnamefont {Y.}~\bibnamefont {Nakamura}},
  \bibinfo {author} {\bibfnamefont {C.~J. P.~M.}\ \bibnamefont {Harmans}}, \
  and\ \bibinfo {author} {\bibfnamefont {J.~E.}\ \bibnamefont {Mooij}},\
  }\href@noop {} {\bibfield  {journal} {\bibinfo  {journal} {Science}\ }\textbf
  {\bibinfo {volume} {299}},\ \bibinfo {pages} {1869} (\bibinfo {year}
  {2003})}\BibitemShut {NoStop}%
\bibitem [{\citenamefont {Neeley}\ \emph {et~al.}(2008)\citenamefont {Neeley},
  \citenamefont {Ansmann}, \citenamefont {Bialczak}, \citenamefont {Hofheinz},
  \citenamefont {Katz}, \citenamefont {Lucero}, \citenamefont {O'Connell},
  \citenamefont {Wang}, \citenamefont {Cleland},\ and\ \citenamefont
  {Martinis}}]{Neeley2008}%
  \BibitemOpen
  \bibfield  {author} {\bibinfo {author} {\bibfnamefont {M.}~\bibnamefont
  {Neeley}}, \bibinfo {author} {\bibfnamefont {M.}~\bibnamefont {Ansmann}},
  \bibinfo {author} {\bibfnamefont {R.~C.}\ \bibnamefont {Bialczak}}, \bibinfo
  {author} {\bibfnamefont {M.}~\bibnamefont {Hofheinz}}, \bibinfo {author}
  {\bibfnamefont {N.}~\bibnamefont {Katz}}, \bibinfo {author} {\bibfnamefont
  {E.}~\bibnamefont {Lucero}}, \bibinfo {author} {\bibfnamefont
  {A.}~\bibnamefont {O'Connell}}, \bibinfo {author} {\bibfnamefont
  {H.}~\bibnamefont {Wang}}, \bibinfo {author} {\bibfnamefont {A.~N.}\
  \bibnamefont {Cleland}}, \ and\ \bibinfo {author} {\bibfnamefont {J.~M.}\
  \bibnamefont {Martinis}},\ }\href {http://dx.doi.org/10.1038/nphys972}
  {\bibfield  {journal} {\bibinfo  {journal} {Nat. Phys.}\ }\textbf {\bibinfo
  {volume} {4}},\ \bibinfo {pages} {523} (\bibinfo {year} {2008})}\BibitemShut
  {NoStop}%
\bibitem [{\citenamefont {Wallraff}\ \emph {et~al.}(2004)\citenamefont
  {Wallraff}, \citenamefont {Schuster}, \citenamefont {Blais}, \citenamefont
  {Frunzio}, \citenamefont {Huang}, \citenamefont {Majer}, \citenamefont
  {Kumar}, \citenamefont {Girvin},\ and\ \citenamefont
  {Schoelkopf}}]{Wallraff2004}%
  \BibitemOpen
  \bibfield  {author} {\bibinfo {author} {\bibfnamefont {A.}~\bibnamefont
  {Wallraff}}, \bibinfo {author} {\bibfnamefont {D.~I.}\ \bibnamefont
  {Schuster}}, \bibinfo {author} {\bibfnamefont {A.}~\bibnamefont {Blais}},
  \bibinfo {author} {\bibfnamefont {L.}~\bibnamefont {Frunzio}}, \bibinfo
  {author} {\bibfnamefont {R.-S.}\ \bibnamefont {Huang}}, \bibinfo {author}
  {\bibfnamefont {J.}~\bibnamefont {Majer}}, \bibinfo {author} {\bibfnamefont
  {S.}~\bibnamefont {Kumar}}, \bibinfo {author} {\bibfnamefont {S.~M.}\
  \bibnamefont {Girvin}}, \ and\ \bibinfo {author} {\bibfnamefont {R.~J.}\
  \bibnamefont {Schoelkopf}},\ }\href {http://dx.doi.org/10.1038/nature02851}
  {\bibfield  {journal} {\bibinfo  {journal} {Nature}\ }\textbf {\bibinfo
  {volume} {431}},\ \bibinfo {pages} {162} (\bibinfo {year}
  {2004})}\BibitemShut {NoStop}%
\bibitem [{\citenamefont {Lupascu}\ \emph {et~al.}(2007)\citenamefont
  {Lupascu}, \citenamefont {Saito}, \citenamefont {Picot}, \citenamefont
  {de~Groot}, \citenamefont {Harmans},\ and\ \citenamefont
  {Mooij}}]{Lupascu2007}%
  \BibitemOpen
  \bibfield  {author} {\bibinfo {author} {\bibfnamefont {A.}~\bibnamefont
  {Lupascu}}, \bibinfo {author} {\bibfnamefont {S.}~\bibnamefont {Saito}},
  \bibinfo {author} {\bibfnamefont {T.}~\bibnamefont {Picot}}, \bibinfo
  {author} {\bibfnamefont {P.~C.}\ \bibnamefont {de~Groot}}, \bibinfo {author}
  {\bibfnamefont {C.~J. P.~M.}\ \bibnamefont {Harmans}}, \ and\ \bibinfo
  {author} {\bibfnamefont {J.~E.}\ \bibnamefont {Mooij}},\ }\href
  {http://dx.doi.org/10.1038/nphys509} {\bibfield  {journal} {\bibinfo
  {journal} {Nat. Phys.}\ }\textbf {\bibinfo {volume} {3}},\ \bibinfo {pages}
  {119} (\bibinfo {year} {2007})}\BibitemShut {NoStop}%
\bibitem [{\citenamefont {He}(2011)}]{Potassium}%
  \BibitemOpen
  \bibfield  {author} {\bibinfo {author} {\bibfnamefont {Y.-L.}\ \bibnamefont
  {He}},\ }\href@noop {} {\bibfield  {journal} {\bibinfo  {journal} {Phys. Rev.
  A}\ }\textbf {\bibinfo {volume} {84}},\ \bibinfo {pages} {053414} (\bibinfo
  {year} {2011})}\BibitemShut {NoStop}%
\bibitem [{\citenamefont {Savenko}\ \emph {et~al.}(2012)\citenamefont
  {Savenko}, \citenamefont {Kibis},\ and\ \citenamefont {Shelykh}}]{Savenko}%
  \BibitemOpen
  \bibfield  {author} {\bibinfo {author} {\bibfnamefont {I.~G.}\ \bibnamefont
  {Savenko}}, \bibinfo {author} {\bibfnamefont {O.~V.}\ \bibnamefont {Kibis}},
  \ and\ \bibinfo {author} {\bibfnamefont {I.~A.}\ \bibnamefont {Shelykh}},\
  }\href@noop {} {\bibfield  {journal} {\bibinfo  {journal} {Phys. Rev. A}\
  }\textbf {\bibinfo {volume} {85}},\ \bibinfo {pages} {053818} (\bibinfo
  {year} {2012})}\BibitemShut {NoStop}%
\bibitem [{\citenamefont {Taminiau}\ \emph {et~al.}(2012)\citenamefont
  {Taminiau}, \citenamefont {Karaveli}, \citenamefont {van Hulst},\ and\
  \citenamefont {Zia}}]{Zia2012}%
  \BibitemOpen
  \bibfield  {author} {\bibinfo {author} {\bibfnamefont {T.~H.}\ \bibnamefont
  {Taminiau}}, \bibinfo {author} {\bibfnamefont {S.}~\bibnamefont {Karaveli}},
  \bibinfo {author} {\bibfnamefont {N.~F.}\ \bibnamefont {van Hulst}}, \ and\
  \bibinfo {author} {\bibfnamefont {R.}~\bibnamefont {Zia}},\ }\href {\doibase
  10.1038/ncomms1984} {\bibfield  {journal} {\bibinfo  {journal} {Nat.
  Commun.}\ }\textbf {\bibinfo {volume} {3}},\ \bibinfo {pages} {979} (\bibinfo
  {year} {2012})}\BibitemShut {NoStop}%
\bibitem [{\citenamefont {Aigouy}\ \emph {et~al.}(2014)\citenamefont {Aigouy},
  \citenamefont {Caze}, \citenamefont {Gredin}, \citenamefont {Mortier},\ and\
  \citenamefont {Carminati}}]{Carminati2014}%
  \BibitemOpen
  \bibfield  {author} {\bibinfo {author} {\bibfnamefont {L.}~\bibnamefont
  {Aigouy}}, \bibinfo {author} {\bibfnamefont {A.}~\bibnamefont {Caze}},
  \bibinfo {author} {\bibfnamefont {P.}~\bibnamefont {Gredin}}, \bibinfo
  {author} {\bibfnamefont {M.}~\bibnamefont {Mortier}}, \ and\ \bibinfo
  {author} {\bibfnamefont {R.}~\bibnamefont {Carminati}},\ }\href@noop {}
  {\bibfield  {journal} {\bibinfo  {journal} {Phys. Rev. Lett.}\ }\textbf
  {\bibinfo {volume} {113}},\ \bibinfo {pages} {076101} (\bibinfo {year}
  {2014})}\BibitemShut {NoStop}%
\bibitem [{\citenamefont {Kasperczyk}\ \emph {et~al.}(2015)\citenamefont
  {Kasperczyk}, \citenamefont {Person}, \citenamefont {Ananias}, \citenamefont
  {Carlos},\ and\ \citenamefont {Novotny}}]{Novotny2015}%
  \BibitemOpen
  \bibfield  {author} {\bibinfo {author} {\bibfnamefont {M.}~\bibnamefont
  {Kasperczyk}}, \bibinfo {author} {\bibfnamefont {S.}~\bibnamefont {Person}},
  \bibinfo {author} {\bibfnamefont {D.}~\bibnamefont {Ananias}}, \bibinfo
  {author} {\bibfnamefont {L.}~\bibnamefont {Carlos}}, \ and\ \bibinfo {author}
  {\bibfnamefont {L.}~\bibnamefont {Novotny}},\ }\href@noop {} {\bibfield
  {journal} {\bibinfo  {journal} {Phys. Rev. Lett.}\ }\textbf {\bibinfo
  {volume} {114}},\ \bibinfo {pages} {163903} (\bibinfo {year}
  {2015})}\BibitemShut {NoStop}%
\bibitem [{\citenamefont {Landau}\ and\ \citenamefont
  {Lifshitz}(1976)}]{Landau}%
  \BibitemOpen
  \bibfield  {author} {\bibinfo {author} {\bibfnamefont {L.~D.}\ \bibnamefont
  {Landau}}\ and\ \bibinfo {author} {\bibfnamefont {E.}~\bibnamefont
  {Lifshitz}},\ }\href
  {http://www.amazon.com/Quantum-Mechanics-Third-Edition-Non-Relativistic/dp/0750635398}
  {\emph {\bibinfo {title} {{Quantum Mechanics}}}}\ (\bibinfo  {publisher}
  {Butterworth-Heinemann},\ \bibinfo {year} {1976})\ p.\ \bibinfo {pages}
  {689}\BibitemShut {NoStop}%
\bibitem [{\citenamefont {Meath}\ and\ \citenamefont {Power}(1984)}]{Meath84}%
  \BibitemOpen
  \bibfield  {author} {\bibinfo {author} {\bibfnamefont {W.~J.}\ \bibnamefont
  {Meath}}\ and\ \bibinfo {author} {\bibfnamefont {E.~A.}\ \bibnamefont
  {Power}},\ }\href {\doibase 10.1080/00268978400100411} {\bibfield  {journal}
  {\bibinfo  {journal} {Molecular Physics}\ }\textbf {\bibinfo {volume} {51}},\
  \bibinfo {pages} {585} (\bibinfo {year} {1984})}\BibitemShut {NoStop}%
\bibitem [{\citenamefont {Metic}\ and\ \citenamefont {Meath}(1985)}]{Meath85}%
  \BibitemOpen
  \bibfield  {author} {\bibinfo {author} {\bibfnamefont {M.~A.}\ \bibnamefont
  {Metic}}\ and\ \bibinfo {author} {\bibfnamefont {W.~J.}\ \bibnamefont
  {Meath}},\ }\href@noop {} {\bibfield  {journal} {\bibinfo  {journal} {Phys.
  Lett.}\ }\textbf {\bibinfo {volume} {108A}},\ \bibinfo {pages} {340}
  (\bibinfo {year} {1985})}\BibitemShut {NoStop}%
\bibitem [{\citenamefont {Bavli}\ \emph {et~al.}(1990)\citenamefont {Bavli},
  \citenamefont {Heller},\ and\ \citenamefont {Band}}]{Band90}%
  \BibitemOpen
  \bibfield  {author} {\bibinfo {author} {\bibfnamefont {R.}~\bibnamefont
  {Bavli}}, \bibinfo {author} {\bibfnamefont {D.~F.}\ \bibnamefont {Heller}}, \
  and\ \bibinfo {author} {\bibfnamefont {Y.~B.}\ \bibnamefont {Band}},\
  }\href@noop {} {\bibfield  {journal} {\bibinfo  {journal} {Phys. Rev. A}\
  }\textbf {\bibinfo {volume} {41}},\ \bibinfo {pages} {3960} (\bibinfo {year}
  {1990})}\BibitemShut {NoStop}%
\bibitem [{\citenamefont {Nakai}\ and\ \citenamefont {Meath}(1992)}]{Meath92}%
  \BibitemOpen
  \bibfield  {author} {\bibinfo {author} {\bibfnamefont {S.}~\bibnamefont
  {Nakai}}\ and\ \bibinfo {author} {\bibfnamefont {W.~J.}\ \bibnamefont
  {Meath}},\ }\href@noop {} {\bibfield  {journal} {\bibinfo  {journal} {J.
  Chem. Phys.}\ }\textbf {\bibinfo {volume} {96}},\ \bibinfo {pages} {4991}
  (\bibinfo {year} {1992})}\BibitemShut {NoStop}%
\bibitem [{\citenamefont {Meath}(2016)}]{AIP}%
  \BibitemOpen
  \bibfield  {author} {\bibinfo {author} {\bibfnamefont {W.~J.}\ \bibnamefont
  {Meath}},\ }\href@noop {} {\bibfield  {journal} {\bibinfo  {journal} {AIP
  Advances}\ }\textbf {\bibinfo {volume} {6}},\ \bibinfo {pages} {075202}
  (\bibinfo {year} {2016})}\BibitemShut {NoStop}%
\bibitem [{\citenamefont {Kibis}\ \emph {et~al.}(2009)\citenamefont {Kibis},
  \citenamefont {Slepyan}, \citenamefont {Maksimenko},\ and\ \citenamefont
  {Hoffmann}}]{Hoffmann}%
  \BibitemOpen
  \bibfield  {author} {\bibinfo {author} {\bibfnamefont {O.~V.}\ \bibnamefont
  {Kibis}}, \bibinfo {author} {\bibfnamefont {G.~Y.}\ \bibnamefont {Slepyan}},
  \bibinfo {author} {\bibfnamefont {S.~A.}\ \bibnamefont {Maksimenko}}, \ and\
  \bibinfo {author} {\bibfnamefont {A.}~\bibnamefont {Hoffmann}},\ }\href@noop
  {} {\bibfield  {journal} {\bibinfo  {journal} {Phys. Rev. Lett.}\ }\textbf
  {\bibinfo {volume} {102}},\ \bibinfo {pages} {023601} (\bibinfo {year}
  {2009})}\BibitemShut {NoStop}%
\bibitem [{\citenamefont {Macovei}\ \emph {et~al.}(2015)\citenamefont
  {Macovei}, \citenamefont {Mishra},\ and\ \citenamefont {Keitel}}]{Macovei}%
  \BibitemOpen
  \bibfield  {author} {\bibinfo {author} {\bibfnamefont {M.}~\bibnamefont
  {Macovei}}, \bibinfo {author} {\bibfnamefont {M.}~\bibnamefont {Mishra}}, \
  and\ \bibinfo {author} {\bibfnamefont {C.~H.}\ \bibnamefont {Keitel}},\
  }\href@noop {} {\bibfield  {journal} {\bibinfo  {journal} {Phys. Rev. A}\
  }\textbf {\bibinfo {volume} {92}},\ \bibinfo {pages} {013846} (\bibinfo
  {year} {2015})}\BibitemShut {NoStop}%
\bibitem [{\citenamefont {Brown}\ \emph {et~al.}(2000)\citenamefont {Brown},
  \citenamefont {Meath},\ and\ \citenamefont {Tran}}]{Meath2000}%
  \BibitemOpen
  \bibfield  {author} {\bibinfo {author} {\bibfnamefont {A.}~\bibnamefont
  {Brown}}, \bibinfo {author} {\bibfnamefont {W.~J.}\ \bibnamefont {Meath}}, \
  and\ \bibinfo {author} {\bibfnamefont {P.}~\bibnamefont {Tran}},\ }\href@noop
  {} {\bibfield  {journal} {\bibinfo  {journal} {Phys. Rev. A}\ }\textbf
  {\bibinfo {volume} {63}},\ \bibinfo {pages} {013403} (\bibinfo {year}
  {2000})}\BibitemShut {NoStop}%
\bibitem [{\citenamefont {Glenn}\ \emph {et~al.}(2013)\citenamefont {Glenn},
  \citenamefont {Limes}, \citenamefont {Pankovich}, \citenamefont {Saam},\ and\
  \citenamefont {Raikh}}]{Raikh}%
  \BibitemOpen
  \bibfield  {author} {\bibinfo {author} {\bibfnamefont {R.}~\bibnamefont
  {Glenn}}, \bibinfo {author} {\bibfnamefont {M.~E.}\ \bibnamefont {Limes}},
  \bibinfo {author} {\bibfnamefont {B.}~\bibnamefont {Pankovich}}, \bibinfo
  {author} {\bibfnamefont {B.}~\bibnamefont {Saam}}, \ and\ \bibinfo {author}
  {\bibfnamefont {M.~E.}\ \bibnamefont {Raikh}},\ }\href@noop {} {\bibfield
  {journal} {\bibinfo  {journal} {Phys. Rev. B}\ }\textbf {\bibinfo {volume}
  {87}},\ \bibinfo {pages} {155128} (\bibinfo {year} {2013})}\BibitemShut
  {NoStop}%
\bibitem [{\citenamefont {Todorov}\ \emph {et~al.}(2010)\citenamefont
  {Todorov}, \citenamefont {Andrews}, \citenamefont {Colombelli}, \citenamefont
  {De~Liberato}, \citenamefont {Ciuti}, \citenamefont {Klang}, \citenamefont
  {Strasser},\ and\ \citenamefont {Sirtori}}]{Sirtori2010USC}%
  \BibitemOpen
  \bibfield  {author} {\bibinfo {author} {\bibfnamefont {Y.}~\bibnamefont
  {Todorov}}, \bibinfo {author} {\bibfnamefont {A.~M.}\ \bibnamefont
  {Andrews}}, \bibinfo {author} {\bibfnamefont {R.}~\bibnamefont {Colombelli}},
  \bibinfo {author} {\bibfnamefont {S.}~\bibnamefont {De~Liberato}}, \bibinfo
  {author} {\bibfnamefont {C.}~\bibnamefont {Ciuti}}, \bibinfo {author}
  {\bibfnamefont {P.}~\bibnamefont {Klang}}, \bibinfo {author} {\bibfnamefont
  {G.}~\bibnamefont {Strasser}}, \ and\ \bibinfo {author} {\bibfnamefont
  {C.}~\bibnamefont {Sirtori}},\ }\href@noop {} {\bibfield  {journal} {\bibinfo
   {journal} {Phys. Rev. Lett.}\ }\textbf {\bibinfo {volume} {105}},\ \bibinfo
  {pages} {196402} (\bibinfo {year} {2010})}\BibitemShut {NoStop}%
\bibitem [{\citenamefont {De~Liberato}(2014)}]{Liberato2014}%
  \BibitemOpen
  \bibfield  {author} {\bibinfo {author} {\bibfnamefont {S.}~\bibnamefont
  {De~Liberato}},\ }\href@noop {} {\bibfield  {journal} {\bibinfo  {journal}
  {Phys. Rev. Lett.}\ }\textbf {\bibinfo {volume} {112}},\ \bibinfo {pages}
  {016401} (\bibinfo {year} {2014})}\BibitemShut {NoStop}%
\bibitem [{\citenamefont {{De Oliveira}}\ \emph {et~al.}(1990)\citenamefont
  {{De Oliveira}}, \citenamefont {Kim}, \citenamefont {Knight},\ and\
  \citenamefont {Buek}}]{DeOliveira1990}%
  \BibitemOpen
  \bibfield  {author} {\bibinfo {author} {\bibfnamefont {F.~A.~M.}\
  \bibnamefont {{De Oliveira}}}, \bibinfo {author} {\bibfnamefont {M.~S.}\
  \bibnamefont {Kim}}, \bibinfo {author} {\bibfnamefont {P.~L.}\ \bibnamefont
  {Knight}}, \ and\ \bibinfo {author} {\bibfnamefont {V.}~\bibnamefont
  {Buek}},\ }\href {\doibase 10.1103/PhysRevA.41.2645} {\bibfield  {journal}
  {\bibinfo  {journal} {Phys. Rev. A}\ }\textbf {\bibinfo {volume} {41}},\
  \bibinfo {pages} {2645} (\bibinfo {year} {1990})}\BibitemShut {NoStop}%
\bibitem [{\citenamefont {Philbin}(2014)}]{Philbin2014}%
  \BibitemOpen
  \bibfield  {author} {\bibinfo {author} {\bibfnamefont {T.}~\bibnamefont
  {Philbin}},\ }\href@noop {} {\bibfield  {journal} {\bibinfo  {journal}
  {‎Am. J. Phys.}\ }\textbf {\bibinfo {volume} {82}},\ \bibinfo {pages} {742}
  (\bibinfo {year} {2014})}\BibitemShut {NoStop}%
\bibitem [{\citenamefont {Braak}(2011)}]{Braak}%
  \BibitemOpen
  \bibfield  {author} {\bibinfo {author} {\bibfnamefont {D.}~\bibnamefont
  {Braak}},\ }\href@noop {} {\bibfield  {journal} {\bibinfo  {journal} {Phys.
  Rev. Lett.}\ }\textbf {\bibinfo {volume} {107}},\ \bibinfo {pages} {100401}
  (\bibinfo {year} {2011})}\BibitemShut {NoStop}%
\bibitem [{\citenamefont {Chen}\ \emph {et~al.}(2012)\citenamefont {Chen},
  \citenamefont {Wang}, \citenamefont {He}, \citenamefont {Liu},\ and\
  \citenamefont {Wang}}]{Chen2012}%
  \BibitemOpen
  \bibfield  {author} {\bibinfo {author} {\bibfnamefont {Q.~H.}\ \bibnamefont
  {Chen}}, \bibinfo {author} {\bibfnamefont {C.}~\bibnamefont {Wang}}, \bibinfo
  {author} {\bibfnamefont {S.}~\bibnamefont {He}}, \bibinfo {author}
  {\bibfnamefont {T.}~\bibnamefont {Liu}}, \ and\ \bibinfo {author}
  {\bibfnamefont {K.~L.}\ \bibnamefont {Wang}},\ }\href@noop {} {\bibfield
  {journal} {\bibinfo  {journal} {Phys. Rev. A}\ }\textbf {\bibinfo {volume}
  {86}},\ \bibinfo {pages} {023822} (\bibinfo {year} {2012})}\BibitemShut
  {NoStop}%
\bibitem [{\citenamefont {Tanas}\ \emph {et~al.}(1992)\citenamefont {Tanas},
  \citenamefont {Murzakhmetov}, \citenamefont {Gantsog},\ and\ \citenamefont
  {Chizhov}}]{Tanas1992}%
  \BibitemOpen
  \bibfield  {author} {\bibinfo {author} {\bibfnamefont {R.}~\bibnamefont
  {Tanas}}, \bibinfo {author} {\bibfnamefont {B.~K.}\ \bibnamefont
  {Murzakhmetov}}, \bibinfo {author} {\bibfnamefont {T.}~\bibnamefont
  {Gantsog}}, \ and\ \bibinfo {author} {\bibfnamefont {A.~V.}\ \bibnamefont
  {Chizhov}},\ }\href@noop {} {\bibfield  {journal} {\bibinfo  {journal}
  {Quantum Opt.}\ }\textbf {\bibinfo {volume} {4}},\ \bibinfo {pages} {1}
  (\bibinfo {year} {1992})}\BibitemShut {NoStop}%
\bibitem [{\citenamefont {Arfken}(1972)}]{Arfken1972}%
  \BibitemOpen
  \bibfield  {author} {\bibinfo {author} {\bibfnamefont {G.}~\bibnamefont
  {Arfken}},\ }\href@noop {} {\emph {\bibinfo {title} {{Mathematical Methods
  for Physicists}}}}\ (\bibinfo  {publisher} {Academic Press},\ \bibinfo {year}
  {1972})\ p.\ \bibinfo {pages} {642}\BibitemShut {NoStop}%
\bibitem [{\citenamefont {Irish}(2007)}]{Irish}%
  \BibitemOpen
  \bibfield  {author} {\bibinfo {author} {\bibfnamefont {E.~K.}\ \bibnamefont
  {Irish}},\ }\href@noop {} {\bibfield  {journal} {\bibinfo  {journal} {Phys.
  Rev. Lett.}\ }\textbf {\bibinfo {volume} {99}},\ \bibinfo {pages} {173601}
  (\bibinfo {year} {2007})}\BibitemShut {NoStop}%
\bibitem [{\citenamefont {Meystre}\ and\ \citenamefont
  {Sargent}(2007)}]{Meystre}%
  \BibitemOpen
  \bibfield  {author} {\bibinfo {author} {\bibfnamefont {P.}~\bibnamefont
  {Meystre}}\ and\ \bibinfo {author} {\bibfnamefont {M.}~\bibnamefont
  {Sargent}},\ }\href@noop {} {\emph {\bibinfo {title} {{Elements of Quantum
  Optics}}}}\ (\bibinfo  {publisher} {Springer-Verlag Berlin Heidelberg},\
  \bibinfo {year} {2007})\BibitemShut {NoStop}%
\end{thebibliography}%
\end{document}